\documentstyle[epsf,11pt]{article}

\newcommand{\rf}[1]{(\ref{#1})}
\newcommand{\bea}{\begin{eqnarray}}
\newcommand{\eea}{\end{eqnarray}}

\newcommand{\e}{{\rm e}}
\renewcommand{\d}{{\rm d}}

\newcommand{\g}{\gamma}
\newcommand{\G}{\Gamma}
\renewcommand{\l}{\lambda}
\renewcommand{\b}{\beta}
\renewcommand{\a}{\alpha}
\newcommand{\n}{\nu}
\newcommand{\m}{\mu}

\newcommand{\th}{\theta}
\newcommand{\del}{\delta}
\newcommand{\Del}{\Delta}
\newcommand{\sg}{\sigma}

\newcommand{\vph}{\varphi}

\newcommand{\oh}{\frac{1}{2}}

\newcommand{\dt}{\triangle}
\newcommand{\hdt}{\hat{\dt}}

\newcommand{\prt}{\partial}
\newcommand{\dph}{\del \phi}
\newcommand{\tph}{{\tilde{\phi}}}
\newcommand{\pcl}{{\phi_{cl}}}

\newcommand{\lnm}{{\l_{Nm}}}

\newcommand{\cD}{{\cal D}}

\newcommand{\cN}{{\cal N}}

\newcommand{\hg}{\hat{g}}
\newcommand{\hR}{\hat{R}}
\newcommand{\hk}{\hat{k}}
\newcommand{\hs}{\hat{s}}
\newcommand{\hn}{\hat{n}}

\newcommand{\no}{\nonumber}
\newcommand{\nn}{\no\\}
\def\void{}
\def\labelmark{}

\newenvironment{formula}[1]{\def\labelname{#1}
\ifx\void\labelname\def\junk{\begin{displaymath}}
\else\def\junk{\begin{equation}\label{\labelname}}\fi\junk}%
{\ifx\void\labelname\def\junk{\end{displaymath}}
\else\def\junk{\end{equation}}\fi\junk\labelmark\def\labelname{}}

{\ifx\void\labelname\def\junk{\end{array}\end{displaymath}}
\else\def\junk{\end{array}\right.\end{equation}}
\fi\junk\labelmark\def\labelname{}\def\junk{}
}

\newcommand{\beq}{\begin{formula}}
\newcommand{\eeq}{\end{formula}}
\newcommand{\beqv}{\begin{formula}{}}

\newcommand{\bg}{\breve{g}}
\newcommand{\bm}[1]{\mbox{\boldmath $#1$}}
\newcommand {\cleqn}{\setcounter{equation}{0}}

\begin{document}
\topmargin 0pt
\oddsidemargin 5mm
\headheight 0pt
\headsep 0pt
\topskip 9mm

\hfill    NBI-HE-97-04

\hfill    EPHOU-97-002

\hfill hep-th/9702019

\hfill February 1997

\begin{center}
\vspace{24pt}
{ \large \bf Non-critical open strings 
beyond the semi-classical approximation}

\vspace{24pt}

{\sl J. Ambj\o rn}\footnote{email ambjorn@nbivms.nbi.dk} \\
 The Niels Bohr Institute\\
Blegdamsvej 17, DK-2100 Copenhagen \O , Denmark, \\

\vspace{12pt}
  
{\sl K. Hayasaka}\footnote{email hayasaka@particle.phys.hokudai.ac.jp} 
and 
{\sl R. Nakayama}\footnote{email nakayama@particle.phys.hokudai.ac.jp}
\\
Division of Physics, Graduate School of Science,\\
Hokkaido University, Sapporo 060, Japan

\end{center}
\vspace{24pt}

\vfill

\begin{center}
{\bf Abstract}
\end{center}

\vspace{12pt}

\noindent
We studied the lowest order quantum corrections to the macroscopic wave 
functions $\Gamma (A,\ell)$ of non-critical string theory using the 
semi-classical expansion of Liouville theory. By carefully taking the 
perimeter constraint into account  we obtained a new type of boundary 
condition for the Liouville field which is compatible with the 
reparametrization invariance of the boundary and which is not only a mixture 
of Dirichlet and Neumann types but also involves an integral of an 
exponential of the Liouville field along the boundary. This condition 
contains an unknown function of $A/\ell^2$.  
We determined this function by computing part of the one-loop 
corrections to $\Gamma (A,\ell)$.

\vfill

\newpage

\section{Introduction}
\cleqn

The discretized approach  to two-dimensional quantum gravity 
has been very successful. The underlying lattice regularization,
known as dynamical triangulation, provides a rigorous definition
of two-dimensional quantum gravity, as well as two-dimensional 
quantum gravity coupled to a variety of matter fields
\cite{adf,david,kkm}. Somewhat 
surprising, it seems  easier to perform analytic calculations 
in this formalism than in the corresponding continuum formalism,
i.e. Liouville theory coupled to matter fields. The success of 
{\em KPZ}\cite{kpz} and {\em DDK}\cite{dkd} in calculating the critical 
exponents of the conformal field theories  coupled to two-dimensional 
quantum gravity has over-shaded the fact that these quantities are the only 
ones with a physical interpretation which can be calculated in the context 
of Liouville theory.   Actually it has been  possible to calculate 2, 3 and 
4-point functions\cite{gl} and there is also an attempt to obtain two-loop 
amplitudes in the proper-time gauge\cite{rn}. However, extension to 
general loop amplitudes has not been achieved.  Contrary, in the discretized
approach it has been possible to calculate macroscopic loop functions 
\cite{ajm,mss,ackm} as well 
as so-called two-point functions depending on invariant distances
\cite{kkmw,aw}. 
The various techniques used in the discretized approach,
such as matrix models calculations, loop equations or transfer matrices  are
all of combinatorial nature, and the power of these methods 
reflects that the solution of discretized two-dimensional quantum gravity
is of purely combinatorial nature, maybe due to the lack of gravitons.  
The {\it semi-classical limit} of Liouville theory {\it combined} with 
{\it KPZ} scaling has taught us many of the same lessons 
as the discretized approach, again indicating the simplicity of 
two-dimensional gravity.

Nevertheless, it is not a very satisfactory situation that 
observables like the Hartle-Hawking wave-function 
which have a direct generalization to higher dimensional quantum gravity
cannot be calculated using continuum field theory in a straight-forward way. 
One of the purposes of the study of two-dimensional quantum gravity is to 
gain experience in the calculation of reparametrization invariant
observables in a continuum framework. Although generalizations exist of 
the formalism of dynamical triangulations to higher dimensions
\cite{aj,am,ajk} it 
is presently not clear that the discretized approach is a viable 
route to four-dimensional quantum gravity, and it is anyway 
highly desirable to be able to calculate entirely in a 
continuum framework the reparameterization invariant
observables known from the discretized approach. 
As a first step we calculate in the present  paper the lowest orders of 
quantum corrections to macroscopic loop functions in Liouville theory.

\section{The semi-classical limit}
\cleqn

The starting point of the semi-classical expansion is the partition 
function for two-dimensional gravity coupled to a conformal field theory 
with central charge $c <1$.  We will be interested in the particular situation 
where the universe has one boundary, or, alternatively, in the partition 
function for an {\em open} non-critical string theory.\footnote{
A semiclasical analysis of random surfaces with topology of a sphere 
was performed in \cite{zam}. Semi-classical limit of open non-critical 
string theory was also studied in \cite{pm}.} 
After integration over the matter fields we obtain the following partition 
function 
\beq{*1}
Z = \int [\cD \phi]_{\hg} \, \e^{-S_L(\phi,\hg)},
\eeq
where $S_L(\phi,\hg)$ is the Liouville action on a surface $M$  
with boundary $\prt M$\cite{dop}:
\bea
S_L(\phi,\hg) &=& \frac{1}{8\pi} 
\left[\int_M \d^2\xi \sqrt{\hg} \,{\hg}^{ab}\prt_a \phi \prt_b \phi + 
Q \int_M  \d^2\xi \sqrt{\hg} \, \hR \phi + Q \oint_{\prt M} \d\hs\, \hk \phi 
\right] \nn
&&\;\;+ \left[\n \oint_{\prt M} \d s \, k -\n \oint_{\prt M} \d \hat{s} \, 
\hk + t_0 \int_M \d^2\xi \sqrt{\hg} \, 
\e^{\a \phi} +\zeta_0  \oint_{\prt M} \d\hs \, \e^{\oh \a\phi}\right].
\label{*2}
\eea
In this formula we have used the standard notation
\beq{*3}
g_{ab}= \e^{\frac{2}{Q}\, \phi} \hg_{ab},~~~Q=\sqrt{\frac{25-c}{3}},~~~
\a= \oh ( Q-\sqrt{Q^2-8}),
\eeq
where  $\phi$ is the Liouville field, $\hg_{ab}$ is a fixed reference  
metric  with scalar curvature $\hR$ and extrinsic curvature 
$\hk$ at the boundary. Finally, $t_0$ and $\zeta_0$ denote
the bulk and boundary cosmological coupling constants, respectively.
The cosmological terms are needed
as short distance counter terms due to short distance singularities.
The extrinsic curvature term is not associated with divergences, but 
is in general needed since the Gauss--Bonnet theorem, 
\beq{*4}
 \int_M \d^2\xi \sqrt{g} \, R 
+ \oint_{\prt M} \d s \, k  = 4 \pi \chi(M),
\eeq
fixes only the sum of the intrinsic and extrinsic curvature terms.
 From the definition $k n^b = -2t^a \nabla_a t^b$, where $t^a$ and 
$n^a$ are the tangent and outwards unit normal to the boundary 
and $\nabla_a$ the covariant derivative, we obtain
\beq{*4a}
k = \e^{-\frac{\phi}{Q}}\Bigl(\hk + \frac{2}{Q} \prt_{\hn} \phi\Bigr),
\eeq
where $\prt_{\hn} = \hn^a \prt_a$. (Here the definition of $k$ is different 
from that in the first Ref of \cite{dop} by a factor $-2$.)

Eq.\ \rf{*4a} is analogous to the relation 
between $\hR$ and $R$ given by
\beq{*4b}
R = \e^{-\frac{2}{Q} \,\phi}\, \Bigl(\hR -\frac{2}{Q}\hat{\triangle} 
\phi\Bigr).
\eeq
In \rf{*4b} we have used the following notation
\beq{*4d}
\hat{\triangle} = 
\frac{1}{\sqrt{\hg}}\; \prt_a \,\sqrt{\hg}\, \hg^{ab}\prt_b
= \hg^{ab} \hat{\nabla}_a \hat{\nabla}_b,
\eeq
where $\hat{\nabla}_a$ is the covariant derivative with respect to 
the metric $\hg_{ab}$. Eq.\ \rf{*4a} implies 
\beq{*4c}
\oint_{\prt M} \d s \, k = \frac{2}{Q}\oint_{\prt M} \d \hs\, \prt_{\hn} \phi 
+ \oint_{\prt M} \d \hs \, \hk.
\eeq
The last term is just a constant and will play no role.

In the following we will for simplicity assume that $M$ has the topology 
of the disk, since we expect all the additional complications associated 
with the macroscopic boundary to be present independent of the topology
of the ``bulk''. Given this  topology of $M$ we can choose the 
reference metric as
\beq{*5}
\hg_{ab} = \del_{ab},~~~~{\rm i.e.}~~~\hR =0,~~~\hk =2.
\eeq

The partition function \rf{*1} with the action \rf{*2} is
a consequence of the conformal anomaly, {\em and} the {\em DDK}
ansatz which allows us to treat the functional measure for 
$\phi$ as a Gaussian measure for an ordinary scalar field.
Since $\hg$ {\it is} fiducial,
the requirement that $Z$ should be independent of $\hg$ implies 
the relation between $\a$ and $Q$ given in \rf{*3}. 

The starting point for a semi-classical expansion is the solution 
$\pcl$ of the classical equations of motion. One then makes a formal 
expansion in large $Q$, i.e. we write
\beq{*5b}
\phi = \pcl + \tph,
\eeq
where $\pcl$ is of order $Q$ and $\tph$ is of order 1.
The classical equation of motion corresponding to the action 
\rf{*2}  is 
\beq{*5c}
R(g) = -4\pi t_0,
\eeq
where $R(g)$ is the curvature given by \rf{*4b}.

For a positive cosmological constant $t_0$ (which we assume) 
there is no solution if the topology is that of the disk.
However, if we impose the restriction of a constant area
it is possible to perform a semi-classical expansion. Thus
we  write the partition function as
\beq{*6a}
Z(t_0, \zeta_0) = \int_0^\infty \d A\, \e^{-t_0 A} 
\int_0^\infty \d \ell \,\e^{-\zeta_0 \ell }\; \G(A,\ell).
\eeq 
In formula \rf{*6a} $ \ell \G(A,\ell)$ denotes the disk amplitude 
$\omega(A,\ell)$ as usually calculated by matrix 
models\footnote{The factor $\ell$ is present since the matrix 
models disk amplitude is usually calculated for a boundary 
with a marked point.} for  
fixed area and fixed length of the boundary. The path integral 
which defines $\G(A,\ell)$ includes the integration over all
metrics on the disk with fixed area $A$ and fixed boundary 
length $\ell$.  

If we implement the area and perimeter constraints
\bea
&& \int_M \d^2 \xi \sqrt{\hg} \;\e^{\a\phi}= A, \label{Acon}\\
&& \oint_{\partial M} \d\hs \; \e^{\oh \a \phi} = \ell \label{Lcon}
\eea
via Lagrange multipliers $\m_1$ and $\m_2$ 
the action  \rf{*2} is  changed to a new action $S_L(\phi,\m_2,\m_1)$
by the replacement 
\beq{*6b}
t_0 \int_M \d^2 \xi \sqrt{\hg} \, \e^{\a\phi} \to
\m_2 ( \int_M \d^2 \xi \sqrt{\hg} \, \e^{\a\phi}-A)
\eeq
\beq{*6c}
\zeta_0 \oint_{\prt M} \d \hs  \, \e^{\oh \a\phi} \to 
\m_1 ( \oint_{\prt M} \d \hs  \, \e^{\oh \a\phi} -\ell).
\eeq

The starting point of a semi-classical expansion is a stationary 
point of the action. Under a variation $\phi \to \phi+\del \phi$ 
the action changes as follows
\bea
\del S_L(\phi,\m_2,\m_1)& = &
\frac{1}{4 \pi} \int_M \d^2  \xi\, \prt_a \phi \prt_a \dph 
+ \m_2 \a\int_{M} \d^2\xi \, \e^{\a \phi} \dph\nn
&&+ \frac{Q}{4\pi} \oint_{\prt M} \d \hs \, \dph
+ \oh \m_1 \a \oint_{\prt M} \d \hs \, \e^{\oh \a\phi}\, \del \phi
+\n \a \oint_{\prt M} \d \hs\; \prt_{\hn} \dph \nn
&=& \frac{1}{4 \pi} \int_M \d^2  \xi\, \del \phi (-\prt_a^2\phi + 
4\pi\m_2 \a \e^{\a\phi})  \nn
&& +\, \frac{1}{4 \pi} \oint_{\prt M} \d \hs \, 
\del \phi(\prt_{\hn} \phi + Q + 2 \pi \m_1 \a \e^{\oh \a \phi}) \nn
&& + \,\n \a \oint_{\prt M} \d \hs \, \prt_{\hn} \del \phi.
\label{*6}
\eea

Let us first consider the case where $\n =0$.
$\del S_L(\phi,\m_2,\m_1)=0$ requires that  $\phi$ inside $M$ 
is a  solution to  the Liouville equation:
\beq{Liouville}
\prt^2_a \phi = 4\pi \m_2 \a \e^{\a\phi}.
\eeq
On the boundary $\partial M$ we {\em cannot} 
impose the Dirichlet condition on 
$\phi$, since $\phi$ (being related to $\det \, g_{ab}$) 
is not a scalar field. The Dirichlet boundary 
condition imposed on $\phi$  breaks the  
invariance under reparametrizations of the boundary, and this 
invariance is necessary for being able to choose the conformal gauge.  
In order to allow $\del \phi \neq 0$ at the boundary, we have to impose  
\beq{boundary}
\prt_{\hn} \phi +Q + 2\pi \m_1 \a \e^{\oh\a\phi}  = 0 \qquad \mbox{on} \quad 
\partial M.
\eeq
This mixed type condition \rf{boundary} keeps the reparametrization symmetry 
of the boundary intact. 
Finally the variations with respect to the Lagrange multipliers 
enforce the constraints \rf{Acon} and \rf{Lcon}.

A solution up to coordinate transformation is given by
\beq{*7a}
\phi_{cl}(\xi) = \frac{1}{\a} \ln \frac{y}{(1-x \xi^2)^2},
\eeq
where $(\xi_1,\xi_2)$ are Cartesian coordinates of the unit disk and 
where we have used the following parameterization\cite{mss}:
\beq{*7b}
x=1-\frac{4\pi A}{\ell^2} <1~~~~~~~~~
y = 4\frac{A^2}{\ell^2} = \frac{A}{\pi}  (1-x).
\eeq
The area $A$ and the perimeter $\ell$ are related to
the values of $\m_1$ and $\m_2$ by
\beq{*7g}
\m_2(x,y)= \frac{2}{\pi \a^2} \, \frac{x}{y},~~~~~
\m_1(x,y)  = \frac{1}{2\pi \a \sqrt{y}} (-Q(1+x) + 2\a x).
\eeq
The action of this classical solution  (with general $\n$) can easily be 
calculated and we obtain 
\beq{*value}
S_L(\pcl) = \Bigl( \frac{Q}{\a} + \frac{16\pi \n}{Q \a} -1\Bigr) 
\Bigl( \frac{1}{1-x}-1\Bigr)+
\frac{Q}{2\a} \log y - \log (1-x),
\eeq
and in terms of the original variables 
$\ell$ and $A$ we can write
\beq{*value1}
\e^{-S_L (\phi)} = {\rm const.}\, A^{-\frac{Q}{\a}+1}\ell^{\frac{Q}{\a}-2}
\e^{-\frac{1}{4\pi} (\frac{Q}{\a} + \frac{16\pi \n}{Q \a} -1) 
\frac{\ell^2}{A}}.
\eeq

The functional integral has to be performed under the constraint \rf{Acon},
\rf{Lcon}.
Thus it is natural to consider a more general boundary condition of the form
\beq{*bounC1}
\prt_{\hn} \phi +\oh \hk Q  + 2\pi \a \m_1(x,y) \, \e^{\oh \a \phi} = 
\frac{1}{\ell} B(x)
\Bigl( \oint_{\partial M} d \hs e^{\oh \a \phi}-\ell \Bigl) 
~~~~{\rm on}~~~\prt M,
\eeq
where $\hk =2$ for our choice \rf{*5} of background metric. 
This boundary condition is of a mixed type and also involves an integral of 
$e^{\oh \a \phi}$ along the boundary.  We note that \rf{*bounC1} is invariant 
under the transformation 
\beq{*new0}
\hg_{ab} \to \e^\sg \hg_{ab},~~~~\phi \to \phi -\frac{Q}{2}\sg,
\eeq 
provided $\a$ is given by \rf{*3}, because $\e^{\oh \a \phi}$ has, like the 
bulk cosmological term, an extra $\hg$ dependence due to the 
reparametrization-invariant regularization and it is transformed in the 
same way as a normal ordered operator in quantum Liouville theory. 
The right-hand side of \rf{*bounC1} vanishes due to \rf{Lcon}.
Nonetheless this generalization will yield non-trivial results 
because a different choice of $B(x)$ defines a different integration 
measure $\cD \phi$.   

We note that \rf{*bounC1} and \rf{*4c} determine an integral of $k$ 
completely.
\beq{*k}
\oint_{\partial M} ds k = -4\pi \Bigl( 1-\frac{2\a}{Q} \Bigr) + \frac{2\ell^2}
{A} \Bigl( 1-\frac{\a}{Q} \Bigr)
\eeq
The Gauss-Bonnet theorem \rf{*4} then gives
\beq{*R}
\int_M d^2 \xi \sqrt{g} R = \Big( 1- \frac{\a}{Q} \Bigr) \Bigl( 8\pi -
\frac{2\ell^2}{A} \Bigl),
\eeq
which does {\em not} approach the spherical value $8\pi$ 
in the limit $\ell \to 0$.  This indicates that the boundary and bulk 
cosmological terms may be renormalized by finite factors $1-(\a/Q)$ and 
$(1-(\a/Q))^2$, respectively, in addition to the infinite, $\hg$-dependent 
factors.  This point, however, will not affect the results  in this paper. 

 For the boundary condition \rf{*bounC1} it follows
that $S_L$ in \rf{*6} is stationary for any variation 
$\del \phi$ with respects the constraints 
\rf{Acon} and \rf{Lcon}, even if $\n \neq 0$, the reason being that 
to lowest order in $\del \phi$ we have
\beq{Cxy1} 
\prt_{\hn} \del \phi = -\pi \a^2 \m_1(x,y) e^{\oh \a \phi} \; \del \phi
\eeq
and consequently the last two terms in \rf{*6} will vanish due to the 
constraint \rf{Lcon} which implies that 
\beq{const}
\oint_{\prt M} d \hs e^{\oh \a \phi} \, \del \phi=0
\eeq
to first order in $\del \phi$. 
 
Now the quantum fluctuation $\tph$ will satisfy the boundary condition
\beq{btph1}
\prt_{\hn} \tph = \Bigl( \frac{2(1+x)}{\a (1-x)} + \a \Bigr) \Bigl(  e^{\oh 
\a \tph} -1 \Bigr) + B(x) \oint_{\partial M} \d \hs  \Bigr(  e^{\oh\a \tph} -1
 \Bigr)  \qquad \mbox{at} \quad \partial M,
\eeq
and to lowest order of expansion in $1/Q$ \rf{btph1} reads 
\beq{btph}
\partial_{\hn} \tph = \frac{1+x}{1-x} \tph + \oh \a B(x) \oint_{\partial M} d 
\hs \tph.
\eeq
By performing a decomposition of $\tph$ into modes $\tph_0 = \oint(d\hs/2\pi)
\tph$ which are constant at the boundary and $\tph_1$ which satisfy  
$\oint d\hs \tph_1=0$, and by defining a new constant $E(x)$ by 
\beq{C0E}
B(x) = -\frac{1}{\pi \a} \Bigl( \frac{1+x}{1-x} +\frac{2E(x)}{(1-x)^2} \Bigr),
\eeq
we finally write the boundary condition as 
\beq{*Bound1}
\prt_{\hn} \tph_1 = \frac{1+x}{1-x} \tph_1,
\eeq
\beq{*Bound2}
\prt_{\hn} \tph_0 = \frac{-2 E(x)}{(1-x)^2} \tph_0.
\eeq

\rf{*Bound1} and \rf{*Bound2} are  the most general boundary conditions 
compatible with invariance under reparametrizations of the boundary. 
In Appendix A we will show that $E(x)$ must have an expansion
\beq{*Bound3}
E(x) = E_{-1}(1-x)+ E_0 + \frac{E_1}{1-x} + \cdots,
\eeq
If we set $B(x)=0$ in \rf{*bounC1}, we will have $E(x) = -\oh (1-x^2)$ and
this will not satisfy the above criterion. 
In later sections we will discuss how to determine $E(x)$.  

\section{The one-loop correction}
\cleqn
\subsection{The eigenvalues of the quadratic action}

According to the discussion in the last section we have to 
perform the following functional integral
\beq{*10}
Z_{\rm 1-loop} (A,\ell) = 
\int \frac{[\cD \tph]_{\hat{g}}}{Vol_{\hg} PSL(2,\bm{R})}  \; \e^{-S^{(2)}_L (\tph)} \; \del(\int_M \d^2\xi \,
\e^{\a\pcl}\, \tph)\; \del(\int \d \hs \, \e^{\oh \a \pcl}\, \tph)  ,
\eeq
where the quadratic action is given by 
\beq{*11}
S^{(2)}(\tph)= \frac{1}{8\pi} \int_M \d^2\xi \, 
\tph \Bigl[-\prt_a^2 + \frac{8x}{(1-x\xi^2)^2} \Bigr] \tph
\eeq
and the boundary conditions are given by \rf{*Bound1}-\rf{*Bound2}. 
In the conformal gauge on a disk there is a residual gauge symmetry given by 
$PSL(2,\bm{R})$ and the group of isometries is its subgroup 
$SO(2)$. Therefore there are two zero modes and to avoid over-counting we 
have to divide the functional measure by the  volume of this group. 

In order to calculate the eigenvalues of the second order differential 
operator it is convenient to change to spherical coordinates, since the 
classical solution for $x <0$ describes a spherical cap. We have 
\beq{*12}
\xi^1 \pm i \xi^2 = \frac{1}{\sqrt{-x}} \tan \frac{\th}{2}\, \e^{\pm i\vph}
\eeq
where 
\beq{*14}
0 \leq \vph < 2\pi,~~~~~0 \leq \th \leq {\rm Tan}^{-1}\sqrt{-x} \equiv \th_{M}
\eeq
In terms of the spherical variable we have
\beq{*15}
S^{(2)}(\tph) = \frac{1}{8\pi} \int_0^{2\pi} \d \vph \int_0^{\th_M} 
\sin\th\d \th \; \tph \Bigl[ -\hdt -2 \Bigr] \tph ,
\eeq
where $\hdt$ is the Laplacian on the spherical cap \rf{*12}-\rf{*14}:
\beq{*16}
\hdt \equiv \frac{1}{\sin \th} \frac{\prt}{\prt \th} \sin \th 
\frac{\prt}{\prt \th}+\frac{1}{\sin^2 \th} \frac{\prt^2}{\prt \vph^2}.
\eeq
Thus the functions on the spherical cap which diagonalize 
the quadratic action \rf{*15} are the generalized
spherical harmonics satisfying 
\beq{*a1}
-\hdt \,\tph = \l_{Nm} (\l_{Nm}+1) \,\tph.
\eeq
They are defined in terms of the associated 
Legendre functions:
\beq{*a3}
Y_{Nm}(\th,\vph) = \cN_{Nm} e^{i m \vph} 
P^m_{\l_{Nm}} (\cos \th),
\eeq
where $N=0,1,2,\cdots$, $m=-N,-N+1,\cdots,N$ and 
\beq{*a3a}
P^m_\l(w) = (-1)^m (1-w^2)^{\frac{|m|}{2}} 
\frac{\d^{|m|}  P_\l (w)}{\d w^{|m|}}.
\eeq
For the chosen boundary conditions the eigenvalues $\l_{Nm}$ will coincide
with the ordinary eigenvalues of the sphere in the limit where 
$x \to -\infty$, i.e.\ the limit where $\th_M \to \pi$ and the 
cap becomes a sphere. Consequently, we can label the eigenvalues 
$\l_{Nm}$ as done in \rf{*a1}.
The normalization constants ${\cal N}_{Nm}$ are determined by 
\beq{*a5}
 \int_0^{\th_M} \sin \th \d \th \int_0^{2\pi} \d \vph \;
Y_{N'm'}^* (\th,\vph) Y_{Nm}(\th,\vph) = 8\pi^2\del_{N'N}\del_{m'm}.
\eeq
Finally the boundary conditions \rf{*Bound1}--\rf{*Bound2} can be 
written in terms of the associated Legendre functions  as 
\beq{*a6}
(\l_{Nm}+2)(1+x) P^m_{\l_{Nm}}(\cos \th_M)  =  (\l_{Nm} - |m| +1)
P^m_{\l_{Nm}-1}(\cos \th_M)
\eeq
for $m \neq 0$ and, for $m=0$, as 
\beq{*a6z}
\left[ \frac{2E(x)}{1-x}+ \l_{N0} (1+x) \right] P_{\l_{N0}}(\cos \th_M)
= \l_{N0} (1-x) P_{\l_{N0}-1}(\cos \th_M). 
\eeq  

In Appendix A we outline how to find the eigenvalues and the 
normalization constants  as an expansion in $1/(1-x)$. 

\subsection{The functional integral}

We use the expansion 
\beq{*30}
\tph = \sum_{N=0}^\infty \sum_{m=-N}^{N} a_{Nm} \, Y_{Nm}(\th,\vph),~~~~
a_{N,-m}=a^*_{Nm},
\eeq
in \rf{*10} and have to express the area and perimeter constraints 
in terms of the mode expansion \rf{*30}: 
\beq{*31}
\int_M \d^2\xi \e^{\a \pcl} \tph = -\frac{\sqrt{2\pi} y}{4x} 
\sum_{N=0}^\infty a_{N0} \cN_{N0}  \int_{\cos \th_M}^1 \ dw \; 
P_{\l_{N0}}(w) \equiv - \frac{\sqrt{2\pi}}{4} \sum_{N=0}^\infty \a_N a_{N0},
\eeq  
\beq{*32}
\int \d \hs \e^{\oh\a \pcl}  \tph =\frac{ 2\pi \sqrt{ y} }{1-x} 
\sum_{N=0}^\infty a_{N0} {\cal N}_{N0} P_{\l_{N0}}(\cos \th_M) \equiv  
2\pi \sum_{N=0}^\infty \b_N a_{N0}.
\eeq 
Using the integration formula for $P_{\l}(w)$ and the relation between  
hyper-geometric functions and associated Legendre polynomials\cite{grad}  
we have the following explicit expressions for the coefficients $\a_N$ and 
$\b_N$:
\bea
\a_N &=& \frac{y}{-x} \cN_{N0} \frac{\sin \theta_M}{\l_{N0} (\l_{N0}+1)} 
P^1_{\l_{N0}}(\cos \theta_M)\nn
& =& \frac{y}{(1-x)^2} \cN_{N0}\, 
F(1-\l_{N0},2+\l_{N_0};2; \frac{-x}{1-x}),           \label{*33}\\
\b_N &=& \frac{\sqrt{y}}{1-x} \cN_{N0} P_{\l_{N0}}(\cos \theta_M) =
 \frac{\sqrt{y}}{1-x} \cN_{N0} \, F(-\l_{N0},1+\l_{N_0};1;\frac{-x}{1-x}).
\label{*34}
\eea
 
A shift of variables from $\tph$ to $\{a_{Nm}\}$ transforms  
the one-loop integral \rf{*10} to 
\bea
Z_{\rm 1-loop} (A,\ell)\!\!& =&\!\!  \Bigl(\prod_{N,m} 
\int \d a_{Nm}\Bigr) \; \exp\Bigl\{-\pi \sum_{N=0}^\infty 
\sum_{m=-N}^N 
[\lnm(\lnm+1) -2] \, |a_{Nm}|^2\Bigr\} \nn
&& \times \del(\sum_{N} \a_N a_{N0}) \,\del(\sum_N \b_N a_{N0})
\del(a_{11})\del(a_{1,-1})\; \times {\cal J} \times \Delta_{FP}, \label{*35} 
\eea 
where the last two $\del$-functions are caused by 
the zero modes corresponding to $Y_{1,\pm 1 }$ and $\Delta_{FP}$ is the 
Faddeev-Popov determinant.
Because we adopted the normalization \rf{*a5}, we have to shift from the 
functional measure $[{\cal D} \tph]_{\hat{g}}$ to 
$[{\cal D} \tph]_{\bg}$, where $\bg_{ab} = -\frac{1}{2\pi^2} 
\frac{x}{y} e^{\a \pcl} \hat{g}_{ab}$.
This induces a Weyl anomaly and is the origin of the Jacobian ${\cal J}$.
Calculation of $\Delta_{FP}$ and ${\cal J}$ is outlined in Appendix B.
Except for the $\del$-functions the integrals can be performed and we obtain
\beq{*36}
{\cal J} \times \Delta_{FP} \times \hat{K}  
\times \hat{I},
\eeq
where 
\beq{*40a}
\hat{K} \equiv 
\left[\prod_{N=1}^\infty \mathop{\hspace*{3pt}{\prod}'}^{N}_{m=1}
 \frac{1}{\lnm(\lnm+1)-2} \right],
\eeq
and where $\hat{I}$ is the integral over the $a_{N0}$'s.  
The prime in $\prod'$ means that the factor for $N=m=1$ 
must be excluded from the product. 
We can implement the two $\del$-functions via Lagrange multipliers and since 
the integrations over the $a_{N0}$'s and the Lagrange multipliers are 
Gaussian, we get after a straightforward calculation:
\beq{*37}
\hat{I} = \sqrt{\prod_{N=0}^\infty \frac{1}{\l_{N0}(1+\l_{N0})-2}}
\times \frac{1}{\sqrt{XZ-Y^2}},
\eeq
where 
\beq{*38}
X=\sum_{N=0}^\infty \frac{\a_N^2}{\l_{N0}(1+\l_{N0})-2},~~~
Y =\sum_{N=0}^\infty \frac{\a_N \b_N}{\l_{N0}(1+\l_{N0})-2},~~~ 
Z =\sum_{N=0}^\infty \frac{\b_N^2}{\l_{N0}(1+\l_{N0})-2}.
\eeq

The problem is now reduced to the calculation of the infinite
products and sums which appear in \rf{*36}-\rf{*38}. It is a 
two-fold problem. We have to calculate the eigenvalues $\lnm$
and afterwards we have to use these values in the actual calculation
of the infinite sum and products.
It seems  impossible to calculate the eigenvalues  
for a general geometry, but two cases can be treated:
$x=-1$, which corresponds to $\th_M=\pi/2$, i.e.\ to the geometry of 
a bowl, and the limit $x \to -\infty$, i.e.\ the limit where $\ell^2/A
\to 0$. In Appendix A the calculation of the eigenvalues in the limit 
$x \to -\infty$ is sketched and it is shown that $E(x)$ must have an 
expansion \rf{*a6a} in order for the eigenvalues to coincide with those of
the sphere in the limit $x \to - \infty$. We note that the expansions for 
the eigenvalues $\l_{N0}$ contain $\ln (1-x)$. We can also show that 
the expansions for other $\l_{Nm}$ generally contain $\ln(1-x)$ in higher
orders. 

The next problem is the calculation of the infinite products. 
This is again non-trivial from a technical point of view. 
Using the results in Appendix A we obtain after a somewhat lengthy 
calculation the following result:
\begin{eqnarray}
\hat{I} & \propto & \frac{(1-x)^2}{y^{3/2}} \exp \{
\frac{1}{1-x} \{  - (E_{-1} +2) \ln (1-x) 
 -1 - \oh \sum_{N=2}^{\infty} (2N+1) \nonumber \\
& & - (E_{-1}+2) \sum_{N=2}^{\infty} 
\frac{2N+1}{(N-1)(N+2)} \} + O((1-x)^{-2})
\} \label{*39}
\end{eqnarray}
and 
\beq{*40}
\hat{K} \propto  1 + O((1-x)^{-2}).
\eeq
The infinite sums which appear in \rf{*39} may be regularized by the 
zeta-function regularization \cite{zeta}.  However, it turns out that 
infinities get more severe in the next orders of $1/(1-x)$.  This means that 
the infinite product and the expansion in $1/(1-x)$ do not commute. 

Therefore we turned to evaluate $\hat{I}$  by exploiting the zeta-function 
regularization without explicitly computing the eigenvalues.  
Some details of this calculation is provided in Appendix C. 
The results are 
\beq{proN0}
\sqrt{\prod_{N=0}^\infty \frac{1}{\l_{N0}(1+\l_{N0})-2}} =  
\frac{ (1-x)^{5/4} (-x)^{1/8}}{\sqrt{2\sqrt{\pi} (1+x)
\Bigl( E(x) + \frac{2x(1-x)}{1+x} \Bigr) }},
\eeq
\beq{xz-y2}
\frac{1}{\sqrt{XZ-Y^2}} = \frac{(-x)^{3/4} (1-x) \sqrt{3+x^2}}{8\pi y^{3/2} 
\sqrt{\th_M}} 
\left [
\frac{2x}{1+x} \frac{E(x)+\oh (1-x)(3+x)}{E(x)+\frac{2x(1-x)}{1+x}}
+\ln \frac{1}{1-x} \right ]^{-\oh}
\eeq
and the product of the two yields 
\bea
\hat{I} =& & \frac{(-x)^{7/8}(1-x)^{9/4}}{2y^{3/2}} 
\sqrt{\frac{3+x^2}{(4\pi)^{5/2}  \th_M}} 
  \left [ 2x \Bigl(E(x) + \oh (1-x)(3+x) \Bigr) \right. \nonumber \\
& & \left. \qquad + (1+x) \Bigl( E(x) + \frac{2x(1-x)}{1+x} \Bigr)
 \ln \frac{1}{1-x} \right ]^{-\oh}. 
\label{ihat}
\eea

Now all $E(x)$ dependence is contained in \rf{ihat}.  In this 
equation we notice a logarithm $\ln(1-x)$which will cause a factor  
$(\ell^2/A)^{(\ell^2/A)}$ in $Z_{\mbox{1-loop}}$ 
in the limit $\ell^2/A \to 0$. 
On the other hand the matrix model result for the partition  function
for some special background\cite{ajm,mss}
\beq{matrix}
Z_{\mbox{matrix}} = \ell^{\frac{Q}{\a} -3} 
A^{-\frac{Q}{\a}} e^{-\mbox{const.} 
\frac{\ell^2}{A}} 
\eeq
does not contain such a factor and to reproduce the matrix model result 
this logarithm has to be canceled in the product $\hat{I} \times \hat{K}$. 
This requirement will determine $E(x)$ completely.  
Let us notice the following feature of 
\rf{proN0}$\sim$\rf{ihat}.  The logarithm in $\hat{I}$ comes solely from 
\rf{xz-y2}, {\it i.e.}  the area and perimeter constraints, and not from 
\rf{proN0}, {\it i.e.} the product of the eigenvalues. 
This result is somewhat surprising because 
the expansions of $\l_{N0}$ about the limit
$x \to -\infty$ {\em do} contain $\ln(1-x)$'s as shown in Appendix A. 
This observation leads us to conjecture that $\hat{K}$ does not contain 
$\ln(1-x)$, either. Actually, when we compute 
\beq{*441}
\hat{K}_m = \prod_{N=max \{2,m \} }^\infty  \frac{1}{\lnm(\lnm+1)-2}
\eeq
for each $m \geq 1$ by the zeta-function regularization explained in 
Appendix C, we obtain 
\beq{Km}
\hat{K}_1 = \frac{9(1-x)^3}{\pi (-x)^{5/2} (3-x)^2}, \qquad 
\hat{K}_m = \frac{m^2 \{ (m-2)! \}^2}{\pi (1-x) (-x)^{m-\oh}} 
\quad (m \geq 2).
\eeq 
This is a strong argument in favor of the 
absence of $\ln(1-x)$ in  
$\hat{K}=\prod_{m=1}^{\infty} K_m$, although it is of course 
not a rigorous proof. If $\ln(1-x)$ is indeed absent in $\hat{K}$ 
we will have to eliminate $\ln(1-x)$ in \rf{ihat} by choosing
\beq{Edila}
E(x) = \frac{-2x(1-x)}{1+x},
\eeq
and thus the boundary condition has  been  completely determined
by the comparison with the known matrix model results.

\section{Discussion}
\cleqn

In this paper we have 
investigated the macroscopic wave functions of non-critical 
string theory within the framework of the semi-classical expansion.  
We have singled out the most general 
boundary condition for the Liouville field 
that does not break reparametrization invariance of the boundary.
This condition contains one unknown function $E(x)$ of $x=1-4\pi (A/\ell^2)$.
The requirement that terms with $\ln(1-x)$ have to be absent in the 
wave function was used to determine $E(x)$. 
To determine this function we (partially) computed the lowest order quantum 
correction to the wave function.  Then  we determined $E(x)$ to be given 
by \rf{Edila} by calculating the contribution 
$\hat{I}$ to the wave function from the constant modes on the boundary 
$\tph_0$ and by arguing (and partly conjecture) 
that the remaining part, $\hat{K}$, in the 
wave function does not contain $\ln(1-x)$.

The boundary condition \rf{Edila} has interesting property:  
\rf{proN0} becomes infinite when we substitute \rf{Edila}.  
This is due to a new zero mode $Y_{1,0}$ corresponding to the dilatation 
$\xi \to e^{\beta} \xi$ which accidentally appears in the quadratic action 
for this boundary condition.  Even in this case \rf{ihat} remain finite 
because \rf{xz-y2} vanishes. This is quite reasonable  because our system 
does not have dilatation symmetry.  However, it is desirable to be 
able to derive the result \rf{Edila} from  first principles, rather than 
by comparison with matrix model calculations.  The above 
superficial dilatation symmetry may be helpful in this connection.

To complete the one-loop calculation we have to evaluate  $\hat{K}$.  
This work is now in progress and the 
results will be reported elsewhere\cite{ahn}.

\section*{Acknowledgements}

Part of this work was done, while one of the authors (R.~N.) visited the 
Niels Bohr Institute.  He is grateful to the hospitality of the staffs and
the members of the theoretical high energy group there.

The work of R.~N. is supported in part by Grant-in-Aid for Scientific Research
((C)(2) No 07640364), ((A)(1) No 08304024) and Grant-in-Aid for International 
Scientific Research (Joint Research No 07044048) from the Ministry of 
Education, Science and Culture of Japan.

J.A. acknowledges the support of the Professor Visitante Iberdrola
grant and the hospitality at the University of Barcelona, where part
of this work was done.

\appendix

\section{Harmonic Analysis on a Disk}
\cleqn

In this appendix we discuss some aspects of harmonic analysis on the disk.
As described in the text we have to solve the following eigenvalue 
problem on the spherical cap with coordinates $(\th,\vph)$ and 
$\th \in [0,\th_M]$, $0 < \th_M \leq \pi$,  $\cos \th_M = \frac{1+x}{1-x}$:
\beq{*a1z}
-\hdt \,\tph = (\l_{Nm} (\l_{Nm}+1)-2) \,\tph.
\eeq
The boundary condition is specified by performing the decomposition into 
modes $\tph_0$ which are constant at the boundary $\prt M$ and modes
$\tph_1$ which satisfy  $\oint ds \tph_1 =0$. We have from \rf{*Bound2}
\beq{*a2}
\prt_{\hn} \tph_1 = \frac{1+x}{1-x} \tph_1,~~~~~~~~
\prt_{\hn} \tph_0 = \frac{-2E(x)}{(1-x)^2} \tph_0 .
\eeq
In general we image that $E(x)$ in \rf{*Bound2} has an expansion
\beq{*a6a}
E(x) = E_{-1} (1-x) + E_0 + \frac{E_1}{1-x} + \cdots,
\eeq
The special choice \rf{Edila}, 
\beq{*a6b}
E(x) =-\frac{2x(1-x)}{1+x} = -2(1-x) -
2 \sum_{n=0}^{\infty} \frac{2^n}{(1-x)^n}.
\eeq
corresponds to the following equation for $\prt_{\hn} \tph_0$ in \rf{*a2}:
\beq{*a2zz}
\prt_{\hn} \tph_0 =\frac{4x}{1-x^2} \tph_0.
\eeq

\subsection*{Expansion of eigenvalues}
It seems difficult to solve the eigenvalue problem for arbitrary 
$x < 0$. Rather, we try an expansion in powers of $\ell^2/A$, i.e.\
in inverse powers of $1-x$. In this case we have that the 
eigenvalues $\l_{Nm}$ should behave as
\beq{*a7}
\l_{Nm}= N + \sum_{n=m}^\infty \frac{\Del^{(Nm)}_n}{(1-x)^n},
\eeq   
In addition we expect the normalization constants $\cN_{Nm}$ 
of the spherical harmonics \rf{*a3} to have an expansion 
\beq{*a8h}
\cN_{Nm} = 2\sqrt{2}\pi \cN^{(0)}_{Nm} \left(1+ \sum_{n=m}^\infty
\frac{a^{(Nm)}_n}{(1-x)^n} \right),
\eeq
where $\cN^{(0)}_{Nm}$ is the normalization of the ordinary spherical 
harmonics.

In principle the calculation of the corrections $\Del_n^{(Nm)}$ and 
$a_n^{(Nm)}$ is straightforward. In practice it is delicate since 
the associated Legendre functions $P^{m}_{\l_{Nm}}(w)$ are 
singular at $w=-1$ unless $\l_{Nm}$ is an integer and  
 $x \to -\infty$ corresponds precisely to this limit since
$\cos \th_M \to -1$ for $x \to -\infty$. The singular behavior can be 
disentangled by using the representation of 
$P^m_\n (w)$ by hypergeometric functions\cite{grad} :
\beq{*a9}
P^m_\n (w)= (-1)^m 
\frac{\G (\n+m+1)}{m! 2^m \G(1+\n-m)} \; (1-w^2)^{\frac{m}{2}} \;
F\Bigl(m-\n,m+\n+1;m+1;\frac{1-w}{2}\Bigr).
\eeq
In terms of the hypergeometric function the boundary conditions 
read:
\bea
\lefteqn{(m-1)(1-x^2)F\Bigl(\l_{Nm}+m+1,m-\l_{Nm};m+1;\frac{-x}{1-x}\Bigr)}&& \nn
&&~~~~~~~~~~~~~~~~=~
2x F'\Bigl(\l_{Nm}+m+1,m-\l_{Nm};m+1;\frac{-x}{1-x}\Bigr),
\label{*a10}
\eea
\beq{*a11}
E(x) F\Bigl(\l_{N0}+1,-\l_{N0};1;\frac{-x}{1-x})=
x F'(\l_{N0}+1,-\l_{N0};1;\frac{-x}{1-x}\Bigr),
\eeq
where \rf{*a10} is valid for $m >0$. The hypergeometric functions 
in \rf{*a10} and \rf{*a11} become singular when $z=-x/(1-x) \to 1$.
But it is well known how to extend the hypergeometric series 
$F(\a,\b;\g;z)$, defined inside the unit circle $|z| <1$, to that inside the 
region $|1-z| <1$ by Gauss's transformation formula 
(see e.g.\ 9.154 in \cite{grad} for some details). Using the same technique
one can separate the singularities and the cut starting at $z=1$.
After some calculations one obtains for $M=0,1,2,\cdots$ 
\bea
\lefteqn{F(\a,\b;\a+\b-M;z) ~~=} \nn
 && \frac{\G(\a+\b-M)}{\G(\a)\G(\b)} 
\sum_{n=0}^{M-1}(-1)^n 
\frac{(\a-M)_n(\b-M)_n(M-n-1)!}{n!(1-z)^{M-n}}\nn
&& - \frac{(-1)^M}{M!} \frac{\G(\a+\b-M)}{\G(\a-M)\G(\b-M)} 
\Bigl[F(\a,\b;M+1;1-z)\; \ln(1-z) \label{*a12}\\
&& - \sum_{n=0}^\infty \frac{(\a)_n(\b)_n}{(M+1)_n \,n!} (1-z)^n 
\{\psi(n+1)+\psi(n+M+1)-\psi(n+\a)-\psi(n+\b)\}\Bigr].\nonumber
\eea
Here $\psi(z) = \frac{d}{dz} \ln \Gamma (z)$ is the di-gamma function 
(psi function).  In this way we can explicitly control the singularities 
at $z=1$.

By inserting \rf{*a12} in the boundary equations \rf{*a10} and \rf{*a11}
we can show that $\l_{Nm}$ approaches the spherical value $N$, only if $E(x)$
has an expansion \rf{*a6a} and we obtain after a tedious calculation:
\bea
\Del^{(N0)}_1 &=& E_{-1}+ N(N+1), \label{*a13}\\
\Del^{(N0)}_2 &=& (\Del^{(N0)}_1)^2 [ \ln(1-x) + 2(\psi(1+N)-\psi(1))]
\nonumber \\
&& +\Del^{(N0)}_1 (1-2N^2) +E_0 + \oh [N(N+1)]^2, \label{*a15}\\
\Del^{(N1)}_2 &=& \oh (N+2)(N+1)N(N-1), \label{*a16}\\
\Del^{(N2)}_2 &=& -\frac{1}{6} (N+2)(N+1)N (N-1), \label{*a17}
\eea
while all other $\Del^{(Nm)}_{1,2}=0$. 

\subsection*{The normalization of the spherical harmonics}

The normalization is fixed by the convention \rf{*a5}, i.e.\
\beq{*a20}
\Bigl( \cN_{Nm}\Bigr)^{-2} = \frac{1}{4\pi} \int^1_{\frac{1+x}{1-x}} \d w \; 
\Bigl(P^m_\lnm (w)\Bigr)^2 \equiv U_{\l_{Nm}}^m.
\eeq
Using standard recursion relations between $P^m_\l$, $P^{m-1}_\l$ and 
$(P^m_\l)'$ one derives the following  recursion relation between  $U_{\l}^m$:
\bea
U_{\l}^m &=& (\l-m+1)(\l+m) U_{\l}^{m-1}
\nn
&& +  \frac{1}{4\pi}\left(\frac{-4x}{(1-x)^2}\right)^m \; 
P^{m-1}_\l \Bigl(\frac{1+x}{1-x}\Bigr)
P^{m}_\l \Bigl(\frac{1+x}{1-x}\Bigr).\label{*a21}
\eea
This leave us with the task of calculation of $U_{\l}^0$. 

As for the eigenvalues $\l_{Nm}$, 
it seems  difficult  to determine $\cN_{Nm}$ except as an 
expansion in inverse powers of $1-x$. Following \rf{*a8} we write 
\beq{*a8}
\cN_{Nm} = \sqrt{2\pi \frac{(2N+1) (N-m)!}{ (N+m)!}} \left(1+ \sum_{n=m}^\infty
\frac{a^{(Nm)}_n}{(1-x)^n} \right),
\eeq
and want to determine the first coefficients $a^{(Nm)}_n$.
The basic equation for $U_{\l}^0$ becomes
\bea
4\pi U_{\l}^0 & =&
\int^1_{\frac{1+x}{1-x}} \d w \; \Bigl(P_\l (w)\Bigr)^2 
= \int^1_{-1} \d w \Bigl(P_\l (w)\Bigr)^2 - 
 \int^{\frac{1+x}{1-x}}_{-1} \d w \Bigl(P_\l (w)\Bigr)^2 \nn
&=& \frac{\pi^2-2 \sin^2 \pi \l \; \psi'(\l+1)}{\pi^2(\l+\oh)} -
\int^{\frac{1+x}{1-x}}_{-1} dw \left[
F\Bigl(-\l,\l+1;1;\frac{1-w}{2}\Bigr)\right]^2.
\label{*a22}
\eea
We have expressed the associated Legendre function $P_\l(w)$
in terms of a hypergeometric function $F$. The singularities 
at $w=-1$ in the integral \rf{*a22} 
can now be handled by the transformation formula \rf{*a12} 
and the integral calculated by expanding the resulting expression in
powers of $1+w$ near $w=-1$. In this way one obtain, 
using the expressions for the 
eigenvalues $\l_{N0}$, already known from \rf{*a13} and \rf{*a15},
the corrections $a^{(N0)}_n$ to $\cN_{N0}$. We can then apply the 
recursion formula \rf{*a21} to obtain the corrections 
$a^{(Nm)}_n$. Again this involves the singularities of 
$P_\lnm^{m}(w)$ at $w =-1$, and they too have to be dealt with 
using the representation of $P_\lnm^{m}(w)$ 
in terms of hypergeometric functions.
Let us summarize the results of the quite lengthy calculation 
as follows:
\bea
a^{(N0)}_1 &=& \frac{E_{-1}+N(N+1)}{2N+1} + \frac{2N+1}{2},
\label{*a34}\\
a_2^{(N0)} & = &
-\frac{1}{8} (4N^2-2N-3)(2N+1) -(2N+1) \Delta_1^{(N0)} \ln (1-x) \nonumber \\
&& -\Delta_1^{(N0)}\left \{2(2N+1) \left (\psi(1)
-\psi(N+1) \right )+2N-\frac{1}{2} \right \} \nonumber \\
&& +\left (\Delta_1^{(N0)}\right )^2 \left (\psi'(N+1) 
-\frac{1}{2(2N+1)^2} \right ) +\frac{\Delta_2^{(N0)}}{2N+1}, \label{*a35}\\
a^{(N1)}_2 &=& \frac{\Del^{(N1)}_2}{2N+1}, \label{*a36}\\
a^{(N2)}_2&=&  \frac{\Del^{(N2)}_2}{2N+1}, \label{*a36z}
\eea
while all other $a^{(Nm)}_{1,2} =0$.
 
\section{Jacobian ${\cal J}$ and Faddeev-Popov Determinant $\Delta_{FP}$}
\cleqn

\subsection*{Jacobian $\cal J$}
The functional measure $[\cD \tilde{\phi} ]_{\hg}$ is
defined with respect to the norm
\beq{mes1}
\left \| \delta \tilde{\phi} \right \|^2_{\hg} =
\int_M d^2\xi \sqrt{\hg} \left (\delta \tilde{\phi} \right )^2 ,
\eeq
while $[\cD \tilde{\phi} ]_{\bg}$  with respect to
\beq{mes2}
\left \| \delta \tilde{\phi} \right \|^2_{\bg} =
 -\frac{1}{2 \pi^2} \frac{x}{y} \int_M d^2\xi 
e^{\a \phi_{cl}} \sqrt{\hg} \left ( \delta \tilde{\phi} \right )^2 .
\eeq
The Jacobian $\cal J$ defined by
\beq{jac}
\left [ \cD \tilde{\phi} \right ]_{\hg}
= {\cal J} \left [ \cD \tilde{\phi} \right ]_{\bg}
\eeq
is given by Weyl anomaly
\beq{Weyl}
 {\cal J} = \exp \left [ -\frac{1}{3Q^2} S_L \left . \left ( \frac{Q}{2}
\left ( \a \phi_{cl} + \ln \left ( -\frac{1}{2 \pi^2} \frac{x}{y} \right ) 
\right ); \hg  \right ) \right |_{t_0=\xi_0=0} \right ] .
\eeq
By a straightforward calculation we have   
\beq{jac2}
  {\cal J} = \mbox{const.} (-x)^{-\frac{1}{12}} \ 
e^{-\left( \frac{1}{6}+\frac{8\pi \nu}{3Q^2} \right ) \frac{1}{1-x}}.
\eeq

\subsection*{Faddeev--Popov determinant $\Delta_{FP}$}
The residual symmetry in the conformal gauge is given by
\beq{rs}
\delta z=-v+v^* z^2.
\eeq
Here $v$ is an infinitesimal complex parameter.
Under this transformation the variation $\delta \phi$ is given  by 
\beq{rsphi}
\delta \phi = \delta z \prt_z \phi + \frac{Q}{2} \prt_z \delta z
+\mbox{c.c.} 
\eeq
and that of $a_{1,\pm 1}$ by
\beq{vara}
\delta a_{1,1}= (\delta a_{1,-1})^{*} = 
\frac{1}{8\pi^2} \int^{\th_M}_0 \sin \theta d \theta
\int^{2\pi}_0 d\varphi \ \delta \phi \ Y^{*}_{1,1}(\th,\varphi).
\eeq
To leading order in $1/Q$ this gives 
\beq{da}
\delta a_{1,1} =  \frac{v^* Q}{2 \pi}  \cN_{1,1}
\frac{x^2 \sqrt{-x}}{(1-x)^2}
\eeq
and by using 
\beq{N11}
\cN_{1,1} = \frac{\sqrt{3\pi} (1-x)^{3/2}}{-x \sqrt{3-x}},
\eeq
we obtain as the determinant $\partial (a_{1,1}, a_{1,-1})/
\partial (v,v*)$
\beq{FPd}
\Delta_{FP}= \mbox{const.} \left ( \cN_{1,1} \right )^2
\frac{-x^3}{(1-x)^4}=\mbox{const.}\frac{-x}{(1-x)(3-x)}.
\eeq

\section{Zeta-Function Regularization}
\cleqn
\subsection*{Calculation of $\hat{I}$}

Zeta-function regularization\cite{zeta} is known as a powerful method for
computing functional determinants.
Here we will apply the technique in \cite{elizalde} to the calculation of 
$\hat{I}$ via zeta-function regularization.

We first define a function
\beq{z0}
\zeta_0(s) = \sum_{N=0}^\infty 
(\lambda_{N0}-1)^{-s} (\lambda_{N0}+2)^{-s}.
\eeq
Analysis of (\ref{*a6z}) shows that $\lambda_{N0}$ increases linearly
for large $N$ and the series \rf{z0} is defined only for Re $s>\oh$.
We then analytically continue $\zeta_0(s)$ onto the whole complex $s$
plane and afterwards compute the infinite product by
\beq{product}
\prod_{N=0}^\infty \frac{1}{\sqrt{\lambda_{N0}(1+\lambda_{N0})-2}}
= \exp \left [ \oh \zeta'_0(0)  \right ] .
\eeq

As a first step we will rewrite $\zeta_0(s)$ as a contour integral.
\beq{z0-1}
\zeta_0(s)=\int_{\gamma_1}\frac{d \lambda}{2\pi i}
\left [ (\lambda-1) (\lambda+2) \right ]^{-s} \frac{\prt}{\prt \lambda} \ln
G_0(\lambda;x)
\eeq
Here $G_0$ is a function defined by
\beq{G0}
G_0(\lambda;x)=\left [ \frac{2 E(x)}{1-x} + \lambda(1+x) \right ]
P_\lambda(\frac{1+x}{1-x})-\lambda (1-x) P_{\lambda-1}(\frac{1+x}{1-x}) 
\eeq
and the complex $\lambda$ plane is cut along the two cuts $(1,1+i\infty)$,
$(-2,-2-i\infty)$. The contour $\gamma_1$ encircles the poles $\l = \l_{N0} \ \
(N=0,1,2,\cdots)$ counterclockwise. (See fig.1)

By using the identity $P_\lambda(w)=P_{-\lambda-1}(w)$ and the recursion
relation for $P_\l (w)$ we can show 
\beq{G0-}
G_0(-\lambda-1;x)=G_0(\lambda;x) \ \ 
\eeq
and thus the integrand of (\ref{z0-1}) is antisymmetric under a transformation 
$\lambda \rightarrow -\lambda-1$.
Furthermore in the integrand of (\ref{z0-1}) there is no more singularity 
than the  poles at $\lambda=\lambda_{N0},-\lambda_{N0}-1$ 
$\ (N=0,1,2,\cdots)$ and the two cuts.
Therefore we can further rewrite (\ref{z0-1}) as
\beq{z0-2}
\zeta_0(s)=-\int_{\gamma2} \frac{d \lambda}{2\pi i}
\left [ (\lambda-1) (\lambda+2) \right ]^{-s} \frac{\prt}{\prt \lambda}
\ln G_0 (\lambda;x), \ \ 
\eeq
where the new contour $\gamma_2$ encircles the cut $(1,1+i \infty)$
counterclockwise. As long as $\oh <$ Re $s<1$ this integral reduces to that of 
discontinuity along the cut
\beq{z0-3}
\zeta_0(s)=\frac{1}{\pi} e^{\frac{\pi}{2} i s}
\sin \pi s \ \ \int_0^\infty dt \ t^{-s}(3+it)^{-s}
\frac{\prt}{\prt t} \ln G_0(1+it;x). \ \ 
\eeq

In the second step we analytically continue $\zeta_0(s)$ 
to the region Re $s< \oh$ by using an asymptotic expansion of 
$P_\n(\cos \theta)$ for Im $\n \rightarrow +\infty$\cite{AS},
\beq{formula3}
P_{\n} ( \cos \theta )
=\frac{1}{\sqrt{2\pi \sin \theta}} \frac {\G (\n+1)}{\G (\n +3/2)}
e^{-i \left (\n \theta +\frac{\th}{2}-\frac{\pi}{4} \right )}
 \left ( 1+ O \left ( \frac{1}{\n} \right ) \right ) + O \left 
( e^{2i \n \theta} \right ),
\eeq
from which we obtain 
\beq{G0asym}
G_0(\l;x) \sim (x-1)\sqrt{\frac{\l}{2\pi} \sin \theta_M}
e^{-i\left ( \l \theta_M + \frac{1}{2}\th_M-\frac{3\pi}{4} \right )} 
\qquad \qquad \l \rightarrow 1+i \infty. 
\eeq
We decompose $\zeta_0(s)$ into  
\beq{z0a}
\zeta^{(1)}_0(s)=\frac{1}{\pi} e^{\frac{i}{2} \pi s} \sin \pi s \ \ 
\int_0^{\infty} dt \ \ t^{-s}(3+it)^{-s} \frac{\prt}{\prt t}
\left [ \ln G_0(1+it;x)-H_0(1+it;x) \right ]
\eeq
and
\beq{z0b}
\zeta^{(2)}_0(s)=\frac{1}{\pi} e^{\frac{i}{2} \pi s} \sin \pi s \ \ 
\int_0^{\infty} dt \ \ t^{-s}(3+it)^{-s} \frac{\prt}{\prt t}
H_0(1+it;x),
\eeq
where
\beq{H0}
H_0(\l;x)= \ln \left[ (x-1) \sqrt{\frac{\sin \theta_M}{2\pi} } \right]
+\oh \ln \l-i\left ( \l \th_M+\oh \th_M -\frac{3\pi}{4} \right ) \ \ .
\eeq
Now (\ref{z0a}) is well defined for $-\oh <$ Re $s<1$ and we obtain 
\begin{eqnarray}
\zeta^{(1)\prime}_0(0) &=& \int_0^\infty dt \frac{\prt}{\prt t}
\left [ \ln G_0(1+it;x) -H_0(1+it;x) \right ] \\
&=& - \ln G_0(1;x)+H_0(1;x), \label{z0-1d}
\end{eqnarray}
to which we substitute from \rf{G0}  
\beq{lnG0-1}
\ln G_0(1;x)= \ln \left [ \frac{2(1+x)}{(1-x)^2} 
\left \{ E(x)+\frac{2x(1-x)}{1+x} \right \} \right ].
\eeq
To compute (\ref{z0b}) we need an integral 
\beq{In}
{\cal I}_n(s)=\frac{1}{\pi} e^{\frac{i}{2} \pi s}
\sin \pi s \ \
\int_0^\infty dt \ \ t^{-s}(3+it)^{-s}(1+it)^{-n} \ \ .
\eeq
This is valid only for $\frac{1-n}{2} <$ Re $s<1$ but can be analytically
continued using a formula\cite{PBM}
\beq{formula1}
\int_0^\infty dt \ \ t^{\a-1}(t+a)^{-\beta}(t+b)^{-\g}
= b^{-\g}a^{\a-\beta} B(\a,\beta+\g-\a)
F(\a,\g;\beta+\g;1-\frac{a}{b})
\eeq
($0< \mbox{Re } \a < \mbox{Re } (\beta+\g)$,
$| \mbox{arg } a|,|\mbox{arg } b|<\pi$.  $B(\a,\b)$ is Euler's beta function.)
We can easily show 
\beq{In1}
{\cal I}_n(s)= 3^{-s} e^{\frac{\pi i}{2} (2s-1)}
\frac{\G(n+2s-1)}{\G(s)\G(n+s)} F(1-s,s;n+s;\frac{2}{3}),
\eeq
and especially ${\cal I}'_0(0)=\frac{3}{2} i$, ${\cal I}'_1(0)= 
\frac{\pi}{2} $.
Putting \rf{z0a}$\sim$\rf{In} and \rf{In1} together, we obtain
\beq{z0d0}
\zeta_0'(0) = -\ln \left ( E(x) + \frac{2x(1-x)}{1+x} \right ) 
 -\ln \frac{2(1+x)}{(1-x)^3}
+ \oh \ln \frac{\sqrt{-x}}{\pi (1-x)}.
\eeq

Next we define
\beq{Xs}
X(s)=\sum_{N=0}^\infty \left [ ( \l_{N0}-1)(\l_{N0}+2) \right ]^{-s}
\a_N^2,
\eeq
\beq{Ys}
Y(s)=\sum_{N=0}^\infty \left [ ( \l_{N0}-1)(\l_{N0}+2) \right ]^{-s}
\a_N\b_N,
\eeq
\beq{Zs}
Z(s)=\sum_{N=0}^\infty \left [ ( \l_{N0}-1)(\l_{N0}+2) \right ]^{-s}
\b_N^2.
\eeq
After analytic continuation, these functions at $s=1$
will provide regularized values of $X,Y$ and $Z$ defined in \rf{*38}.
To rewrite \rf{Xs} $\sim$ \rf{Zs} as contour integrals we have
to analytically continue $\cN_{N0}$.
This will be done in terms of the relation
\beq{Nl}
\begin{array}[b]{rcl}
\displaystyle
4\pi \left ( \cN_\l \right ) ^{-2}&=& \displaystyle
\int_{\cos \theta_M}^1 dw \left (P_\l(w) \right )^2 \\
&=& \displaystyle
\frac{1}{2\l+1} \left [P_\l( \cos \theta_M ) \ P_{\l-1}( \cos \theta_M )
 + \l P_\l( \cos \theta_M ) \frac{\prt}{\prt \l} \ P_{\l-1}( \cos \theta_M )
\right . \\
&& \displaystyle \left .
-\l P_{\l-1}( \cos \theta_M ) \frac{\prt}{\prt \l} \ P_{\l}( \cos \theta_M )
 -\cos \theta_M \left \{ P_\l ( \cos \theta_M ) \right \}^2 \right ].
\end{array}
\eeq
In the last equality use is made of a formula tabulated in Vol I,
Chapter I\hspace*{-2pt}I\hspace*{-2pt}I, 3.12 of \cite{erde}.
A phase of $\cN_\l$ is chosen to be positive when $\l$ is a real number.
The functions \rf{Xs} $\sim$ \rf{Zs} are now written as
\beq{Xs1}
X(s)=\int_{\gamma_1} \frac{d\l}{2\pi i}  
\left [ ( \l-1)(\l+2) \right ]^{-s}
(\a(\l))^2   \frac{\prt}{\prt \l}  \ln G_0(\l;x) ,
\eeq
\beq{Ys1}
Y(s)=\int_{\gamma_1} \frac{d\l}{2\pi i}  
\left [ ( \l-1)(\l+2) \right ]^{-s}
\a(\l)\b(\l)   \frac{\prt}{\prt \l}  \ln G_0(\l;x) ,
\eeq
\beq{Zs1}
Z(s)=\int_{\gamma_1} \frac{d\l}{2\pi i}  
\left [ ( \l-1)(\l+2) \right ]^{-s}
(\b(\l))^2   \frac{\prt}{\prt \l}  \ln G_0(\l;x) ,
\eeq
where
\beq{al}
\a(\l)=  \frac{y}{-x} \cN_{\l} \frac{\sin \th_M}{\l(\l+1)}
P^1_\l(\cos \th_M), \quad 
\b(\l)= \frac{\sqrt{y}}{1-x} \cN_\l P_\l(\cos \th_M).
\eeq

An asymptotic form of $\cN_\l$ for Im $\l \rightarrow + \infty$ is found
by using \rf{formula3}
\beq{Nl1}
\cN_\l \sim 4\pi \ \l \ 
e^{i\left ( \l + \oh\right )\theta_M +\frac{\pi}{2}i}.
\eeq
By using the recursion formula\cite{grad}
\beq{formula5}
P^1_\l(\cos \theta) = \frac{\l}{\sin \theta} [ \cos \theta P_\l(\cos \theta)
-P_{\l-1}(\cos \theta) ],
\eeq
the following asymptotic forms in the same limit are also obtained.
\beq{ablasymp}
\a(\l) \sim e^{\frac{\pi}{4} i}\frac{2 y}{(1-x) \sqrt{-x}} 
\sqrt{\frac{8\pi}{\l \sin \theta_M}}, \qquad 
\b(\l) \sim e^{\frac{3}{4}\pi i}\frac{\sqrt{ y}}{1-x} 
\sqrt{\frac{8\pi \l}{\sin \theta_M}}.
\eeq

Now the contour $\gamma_1$ can be shifted to $\gamma_2$ as in \rf{z0-2}
and the integrals will be reduced to those of discontinuity along the cut.
Let us first consider
\beq{xs}
X(s)=\frac{\sin \pi s}{\pi} e^{\frac{\pi}{2}is}
\int_0^\infty dt \ \ t^{-s}(3+it)^{-s} ( \a (1+it) )^2 
 \frac{\prt}{\prt t} \ln G_0(1+it;x).
\eeq
This is defined for $0<$ Re $s<1$.
By integrating $X(s)$ by parts  by 
using $t^{-s}=\frac{1}{1-s}\frac{\prt}{\prt t} t^{1-s}$, 
we obtain
\beq{Xs2}
X(s)=\frac{\sin \pi s}{\pi (s-1)} e^{\frac{\pi}{2}is}
\int_0^\infty dt \ \ t^{1-s}\frac{\prt}{\prt t} 
\left \{  (3+it)^{-s} ( \a (1+it) )^2 
 \frac{\prt}{\prt t}  \ln G_0(1+it;x) \right \} .
\eeq
As $ s \rightarrow 1-0$ this yields
\beq{Xs3}
X(1)=\left . \left [ -\frac{1}{3} [\a(\l)]^2  \frac{\prt}{\prt \l} 
\ln G_0(\l;x) \right ] \right |_{\l=1}.
\eeq
Similarly the integral for $Y(s)$ is defined for $\oh <$ Re $s<1$ and we obtain
\beq{Ys3}
 Y(1)=\left . \left [ -\frac{1}{3} \a(\l) \b(\l)   \frac{\prt}{\prt \l} 
\ln G_0(\l;x) \right ] \right |_{\l=1}.
\eeq

Finally $Z(s)$ is slightly complicated.  We first decompose the contour 
integral along $\g_2$ into three terms and then rewrite them as integrals 
along the cut. They are given by
\beq{Z10}
Z^{(1)}(s)=\frac{\sin \pi s}{\pi} e^{\frac{\pi}{2}is} \int_0^\infty
dt \ \ t^{-s}(3+it)^{-s} (\b(1+it))^2 \left [
\frac{\prt}{\prt t} \ln G_0(1+it;x)-\theta_M \right ]          ,
\eeq
\beq{Z20}
Z^{(2)}(s)= \frac{\sin \pi s}{\pi} e^{\frac{\pi}{2}is}  \theta_M \int_0^\infty
 dt \ \ t^{-s}(3+it)^{-s} \left [  (\b(1+it))^2  -(\b(1))^2 \right ]    ,
\eeq
\beq{Z30}
  Z^{(3)}(s)= \frac{\sin \pi s}{\pi} e^{\frac{\pi}{2}is}  \theta_M 
  (\b(1))^2 \int_0^\infty dt \ \ t^{-s}(3+it)^{-s} .
\eeq
The first term, which is valid for $\oh <$ Re $s<1$, can be evaluated at 
$s \to 1-0$ as in \rf{Xs3}.
\beq{Z1}
 Z^{(1)}(1)= \left .\left [ -\frac{1}{3} [ \b(\l) ]^2 \frac{\prt}{\prt \l} \ln
G_0(\l;x) \right ] \right |_{\l=1}
-\frac{i}{3} \theta_M [ \b(1) ]^2 
\eeq
The second one is valid for  $1 <$ Re $s <2$ and  partial integration yields
\bea
Z^{(2)}(s) &=& \frac{\sin \pi s}{\pi(s-1)} e^{\frac{\pi}{2} is}
\th_M \int_0^\infty dt \ \ t^{1-s} \frac{\prt}{\prt t} \left [
(3+it)^{-s} \left \{ (\b(1+it))^2-(\b(1))^2 \right \} \right ] \nonumber \\
& \displaystyle \mathop{\longrightarrow}_{s\rightarrow1+0}   &
-\frac{4\pi y}{(1-x) \sqrt{-x}} \theta_M. 
\label{Z2}
\eea
The third one can be evaluated by using \rf{In},
\beq{Z3}
Z^{(3)}(1) = \th_M \left [\b(1) \right ]^2 {\cal I}_0(1) = \frac{i}{3} \th_M 
\left [\b(1) \right ]^2,
\eeq
which together with \rf{Z1} $\sim$ \rf{Z2} gives 
\beq{Zs3}
Z(1) =  \left .\left [ -\frac{1}{3} [\b(\l)]^2 \frac{\prt}{\prt \l} 
\ln G_0(\l;x) \right ] \right |_{\l=1} -\frac{4\pi y}{(1-x) \sqrt{-x}} 
\theta_M.
\eeq
By combining \rf{Xs3}, \rf{Ys3} and \rf{Zs3} we obtain 
\beq{XZ-Y2}
X(1)Z(1)-Y(1)^2 = \frac{4\pi y}{3(1-x)\sqrt{-x}} \th_M \left [ \a(1) \right ]
^2  \left . \left [  \frac{\prt}{\prt \l} 
\ln G_0(\l;x) \right ] \right |_{\l=1}.
\eeq
Using
\beq{n1}
\cN_1 = \sqrt{\frac{-6\pi(1-x)^3}{x(3+x^2)}}, \quad \a (1) = 
\frac{y \sqrt{48 \pi}}{\sqrt{-x(1-x)(3+x^2)}} 
\eeq
and 
\beq{dellnG0}
\left . \left [ \frac{\prt}{\prt \l} \ln G_0(\l;x) \right ] \right |_{\l=1}
= \frac{2x}{1+x} \frac{E(x)+\oh (1-x)(3+x)}{E(x)+\frac{2x(1-x)}{1+x}}
+\ln \frac{1}{1-x},
\eeq
which follows\cite{grad} from 
\beq{delPl}
\frac{\partial}{\partial \l} P_{\l} (w) |_{\l=1} = w-1+w \ln \frac{1+w}{2},
\eeq
we obtain 
\beq{XZ-Y2f}
XZ-Y^2 =  \frac{64\pi^2 y^3 \th_M}{(-x)^{3/2}(1-x)^2(3+x^2)}  
\left [
\frac{2x}{1+x} \frac{E(x)+\oh (1-x)(3+x)}{E(x)+\frac{2x(1-x)}{1+x}}
+\ln \frac{1}{1-x} \right ].
\eeq
By combining this with \rf{product} and \rf{z0d0} we have 
\bea
\hat{I} =& & \frac{ (-x)^{7/8} (1-x)^{9/4}}{2y^{3/2}} 
\sqrt{\frac{3+x^2}{(4\pi)^{5/2}  \th_M}} 
  \left [ 2x \Bigl(E(x) + \oh (1-x)(3+x) \Bigr) \right. \nonumber \\
& & \left. \qquad + (1+x) \Bigl( E(x) + \frac{2x(1-x)}{1+x} \Bigr)
 \ln \frac{1}{1-x} \right ]^{-\oh}. 
\label{Ihat}
\eea

\subsection*{Zeta function for $\lambda_{Nm} \ ( m \geq 1)$}
We define functions
\beq{zeta1}
\zeta_1(s)=2 \sum_{N=2}^\infty \left ( \lambda_{N1}-1 \right )^{-s}
 \left ( \lambda_{N1}+2 \right )^{-s},
\eeq
\beq{zetas}
\zeta(s)=2 \sum_{m=2}^\infty \sum_{N=m}^\infty 
\left ( \lambda_{Nm}-1 \right )^{-s}
 \left ( \lambda_{Nm}+2 \right )^{-s}
\eeq
and rewrite them as
\beq{zeta1'}
\zeta_1(s) = 2 \int_{\gamma_1} \frac{d\lambda}{2\pi i} 
\left [ (\l-1)(\l+2) \right ]^{-s} \frac{\prt}{\prt \l } 
\ln G_1(\l;x) ,
\eeq
\beq{zetas'}
\zeta(s) = 2 \sum_{m=2}^\infty \int_{\gamma_1} \frac{d\lambda}{2\pi i} 
\left [ (\l-1)(\l+2) \right ]^{-s} \frac{\prt}{\prt \l } 
\ln G_m(\l;x) .
\eeq
The functions $G_m$ are defined by
\beq{G1}
G_1(\l;x)=\frac{\Gamma(\l-1)}{\Gamma(\l+3)}
\left [ (\l-1) \frac{1+x}{1-x}  P^1_{\l} ( \frac{1+x}{1-x} ) -
(\l+1) P^1_{\l-1}( \frac{1+x}{1-x} ) \right ],
\eeq

\beq{Gm}
G_m(\l;x)= (-1)^m \frac{\Gamma(\l-m+1)}{\Gamma(\l+m+1)}
\left [ (\l-1) \frac{1+x}{1-x} P^m_{\l} ( \frac{1+x}{1-x} ) -
(\l+m) P^m_{\l-1}( \frac{1+x}{1-x} ) \right ]
\eeq
\hspace*{\fill}$(m\geq2)$ .\\
The contour $\gamma_1$ encircles counterclockwise $\lambda_{N,1} \ 
(N\geq2)$ and $\lambda_{N,m} \ (N\geq m)$, respectively, as in fig.1.
The contour will then be shifted to $\gamma_2$.

Calculation of $\zeta'_1(0)$ can be performed by the same technique as in the
previous subsection and we obtain
\beq{z1d0}
\zeta'_1(0) = \ln \frac{9(1-x)^3}{\pi(-x)^{\frac{5}{2}}(3-x)^2}.
\eeq
To compute $\zeta'(0)$ we have to subtract from $\ln G_m$ its asymptotic 
expansion
$H_m$ for $\l \rightarrow 1+i\infty$ and $m\rightarrow \infty$ with a
ratio $\frac{\l}{m}$ fixed, up to order $\frac{1}{m}$, and then add this 
expansion to $\ln G_m-H_m$.It turns out that these subtracted terms are more
complicated than those for $\zeta'_0(s)$, $\zeta'_1(s)$ and we have to
perform integration over $\l$ and summation over $m$ before setting $s=0$.
Evaluation of $\zeta'(0) $ is thus left for future investigation.

\section*{Figure Captions}
\begin{description}
\item[Fig.1] The contours $\gamma_1$ and $\gamma_2$ in the complex $\l$ plane.
\end{description}
\epsfxsize=16cm \epsfbox {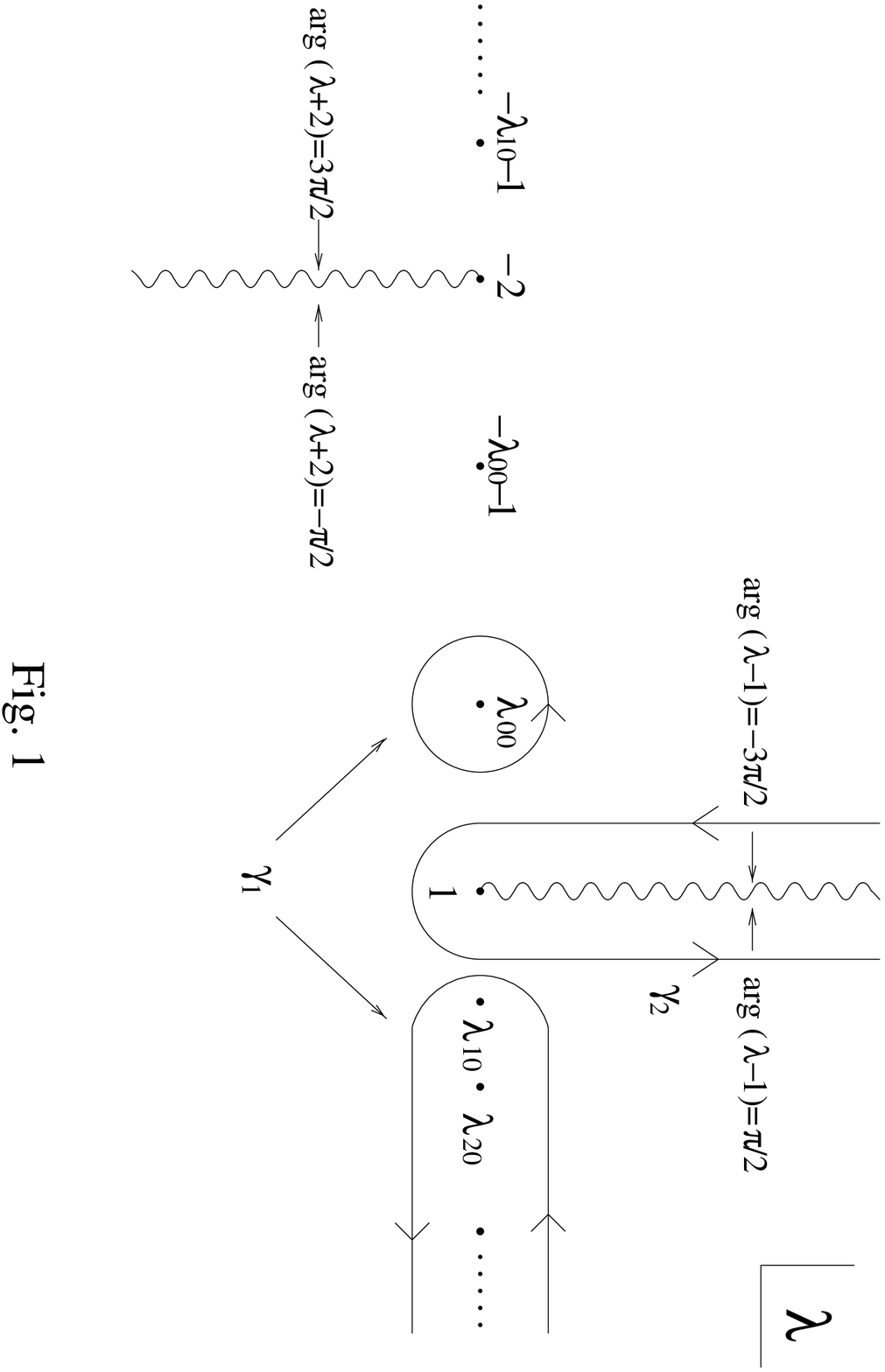}

\end{document}